\documentstyle[aps,prd,epsfig,preprint,tighten]{revtex}
\begin{document}
\draft
\title{			Supernova neutrino oscillations: 
\\
			  A simple analytical approach
}
\author{ 		         G.L.\ Fogli%
$\,^1$,
			         E.\ Lisi%
$\,^1$, 
			        D.\ Montanino%
$\,^2$, 
			       and A.\ Palazzo%
$\,^1$\\[4mm]
}
\address{$^1$Dipartimento di Fisica and Sezione INFN di Bari                \\
             Via Amendola 173, 70126 Bari, Italy                            \\
}
\address{$^2$Dipartimento di Scienza dei Materiali and Sezione INFN di Lecce\\
             Via Arnesano, 73100 Lecce, Italy                               \\ 
}
\maketitle
\begin{abstract}%............................................................
Analyses of observable supernova neutrino oscillation effects require the
calculation of the electron (anti)neutrino survival probability $P_{ee}$ along
a given supernova matter density profile. We propose a simple analytical
prescription for $P_{ee}$, based on a double-exponential form for the crossing
probability and on the concept of maximum violation of adiabaticity. In the
case of two-flavor transitions, the prescription is shown to reproduce
accurately, in the whole neutrino oscillation parameter space, the results of
exact numerical calculations for generic (realistic or power-law) profiles.
The analytical approach is then generalized to cover three-flavor transitions
with (direct or inverse) mass spectrum hierarchy, and to incorporate Earth
matter effects. Compact analytical expressions, explicitly showing the symmetry
properties of $P_{ee}$, are provided for practical calculations.
\end{abstract}%.............................................................
\medskip

%%%%%%%%%%%%%%%%%%%%%%%%%%%%%%%%%%%%%%%%%%%%%%%%%%%%%%%%%%%%%%%%%%%%%%%%%%%%
\section{Introduction}
%%%%%%%%%%%%%%%%%%%%%%%%%%%%%%%%%%%%%%%%%%%%%%%%%%%%%%%%%%%%%%%%%%%%%%%%%%%%

Observable effects of supernova neutrinos in underground detectors represent a
subject of intense investigation in astroparticle physics, both on general
grounds (see the reviews in \cite{Raff,Voge}) and in relation to the SN1987A
event (see \cite{Lo01} for an updated analysis and bibliography). In
particular, flavor oscillations in supernovae may shed light on the problem of
neutrino masses and mixing, by means of the (potentially strong) associated
matter effects (see, e.g., \cite{Raff,KuSN,KuPa,Nuno} for reviews of early
works, and 
\cite{Di00,Lu00,Lu01,Mi01,Du01,Du02,Du03,Ku00,Ka00,Ka01,St01,Ta00,Ta01,Sa01} 
for an incomplete list of recent studies). In particular,  dramatic
effects on oscillations have been predicted, related to the type of neutrino
mass  spectrum hierarchy and to Earth matter crossing.

Given the importance of supernova neutrino oscillations for both particle
physics and astrophysics, it would  be desirable to have a simple and complete
description of the most important quantity involved in the calculations,
namely, of the electron (anti)neutrino survival probability $P_{ee}$.%
%------------
\footnote{The key role of $P_{ee}(\nu)$ and $P_{ee}(\overline\nu)$,  related
to the practical undistinguishability of supernova muon and tau
(anti)neutrinos, is neatly discussed in  \protect\cite{KuPa,Di00}. }
%------------
Approximate treatments of $P_{ee}$ have been proposed in the literature to
cover parts  the parameter space in a piecewise fashion, e.g., by using either
the adiabatic approximation or the so-called resonance condition (see
\cite{Bucc,Di00,Mi01,Du01} for recent examples),  with inherent limitations in
the range of applicability. In particular, it has been recently realized,
first  in the context of solar \cite{Fr01,Li01} and then of supernova
\cite{Ka01,St01}  neutrinos, that the  time-honored resonance condition cannot
be meaningfully extended at large neutrino mixing. Thus, apart from
brute-force numerical calculations of $P_{ee}$ (see, e.g.,
\cite{Mi88,Ta00,Ta01,Sa01}), a truly unified approach, valid in the whole
three-flavor oscillation parameter space and applicable to generic supernova
density profiles, seems still lacking in the literature, as far as we know.

In this work, we propose a simple, unified analytical approach to the
calculation of $P_{ee}$, based on  a double-exponential form for the crossing
probability \cite{Pe88,Kr88} and on the condition of maximum violation of
adiabaticity \cite{Fr01,Li01} (replacing the popular resonance condition). In
the case of two-flavor transitions (Sec.~II), our prescription is shown to
reproduce accurately  the results of exact numerical calculations for generic 
(realistic or power-law) profiles, in the whole neutrino oscillation parameter
space. The analytical approach is then generalized to cover three-flavor
transitions with (direct or inverse) mass spectrum hierarchy (Sec.~III), and to
incorporate Earth matter effects (Sec.~IV). Compact analytical expressions, 
explicitly showing the symmetry properties of $P_{ee}$ and useful for
practical calculations, are  summarized in the final section (Sec.~V), to which
we refer the impatient reader.

%%%%%%%%%%%%%%%%%%%%%%%%%%%%%%%%%%%%%%%%%%%%%%%%%%%%%%%%%%%%%%%%%%%%%%%%%%%%
\section{Two-flavor transitions}
%%%%%%%%%%%%%%%%%%%%%%%%%%%%%%%%%%%%%%%%%%%%%%%%%%%%%%%%%%%%%%%%%%%%%%%%%%%%

In this section we discuss numerical and analytical calculations of the
survival probability $P_{ee}$ for neutrinos and antineutrinos,  assuming
two-family mixing between $\nu_e$ and another active neutrino ($\nu_a=\nu_\mu$
or $\nu_\tau$). A simple analytical prescription will be shown to reproduce
very accurately the exact numerical results for $P_{ee}$.

%%%%%%%%%%%%%%%%%%%%%%%%%%%%%%%%%%%%%%%%%%%%%%%%%%%%%%%%%%%%%%%%%%%%%%%%%%%%
\subsection{$2\nu$ transitions: Notation}

In the case of two-family $\nu_e\to\nu_a$ oscillations ($a=\mu$ or $\tau$), we
label the mass $(m)$ eigenstates $(\nu_1,\,\nu_2)$ so that $\nu_1$ is the
lightest,
%...........................................................................
\begin{equation}
\Delta m^2 = m^2_2-m^2_1>0\ ,
\label{Deltam}
\end{equation}
%...........................................................................
and parametrize the mixing matrix $U$ as
%...........................................................................
\begin{equation}
\left( 
\begin{array}{cc}
U_{e1} & U_{e2} \\
U_{a1} & U_{a2}
\end{array}
\right) =
\left( 
\begin{array}{cc}
 \cos\theta & \sin\theta \\
-\sin\theta & \cos\theta \\
\end{array}
\right)\ ,
\label{U2}
\end{equation}
%...........................................................................
where $\theta\in [0,\,\pi/2]$. In vacuum, $\nu_e\to\nu_a$ oscillations can be
described in terms of the pathlength ($x$) and of the  oscillation wavenumber 
%...........................................................................
\begin{equation}
k=\Delta m^2/2E\ ,
\label{k}
\end{equation}
%...........................................................................
$E$ being the neutrino energy. In matter, the  $\nu_e\to\nu_a$  dynamics 
also  depends on the $\nu_e-\nu_a$ interaction potential  difference
\cite{Matt}
%...........................................................................
\begin{equation}
V(x) = \sqrt{2}\, G_F\, N_e (x)\ ,
\label{V}
\end{equation}
%...........................................................................
where $N_e(x)$ is the electron density profile. In appropriate units,
%...........................................................................
\begin{equation}
\frac{V(x)}{\mathrm{eV}^2/\mathrm{MeV}}= 7.57\times 10^{-8} \,Y_e(x)\,
\frac{\rho(x)}{\mathrm{g}/\mathrm{cm}^3}\ ,
\label{Vu}
\end{equation}
%...........................................................................
where $\rho(x)$ is the matter density and $Y_e(x)$ is the electron/nucleon 
number fraction.

In supernovae, $\rho(x)$ [and thus $V(x)$] can be approximately described by a
power law, $\rho(x)\propto x^{-3}$ \cite{Br82}. In the present work, power-law
potentials  are parametrized as
%...........................................................................
\begin{equation}
V(x) = V_0 \left( \frac{x}{R_\odot}\right)^{-n}\ ,
\label{pow}
\end{equation}
%...........................................................................
where $n=3$ (unless otherwise stated), and distances are conventionally
reported in units of the solar radius, $R_\odot=6.96\times 10^8$~m.

Figure~1 shows an example of ``realistic'' neutrino potential profile  $V(x)$
(dashed curve),  as graphically reduced from the supernova simulation
published in \cite{Sh90} for the case of $M=14.6\, M_\odot$, $M$ being the
mass of the ejecta. In the same figure, the solid line represents the best-fit
power-law potential $(n=3)$, corresponding to take $V_0=1.5\times 10^{-8}$
eV$^2$/MeV in Eq.~(\ref{pow}). For definiteness, we will use the realistic or
the power-law curves in Fig.~1 as representative $V(x)$ profiles for our
calculations. However, our main results are applicable to generic supernova
density profiles.

%%%%%%%%%%%%%%%%%%%%%%%%%%%%%%%%%%%%%%%%%%%%%%%%%%%%%%%%%%%%%%%%%%%%%%%%%%%%
\subsection{$2\nu$ transitions: Neutrinos}

Following \cite{KuPa}, the calculation of $P_{ee}(\nu)$ from the initial $\nu$
state in matter to the final $\nu$ detection in vacuum%
%--------------------------
\footnote{The discussion of possible  Earth matter effects is postponed to
Sec.~IV.}
%------------------------- 
can be factorized as
%...........................................................................
\begin{equation}
P_{ee}(\nu) =
\left(
\begin{array}{cc}
1\;,& 0
\end{array}
\right)
\left(
\begin{array}{cc}
\cos^2\theta & \sin^2\theta \\
\sin^2\theta & \cos^2\theta 
\end{array}
\right)
\left(
\begin{array}{cc}
1-P_c(\nu) & P_c(\nu) \\
P_c(\nu) & 1-P_c(\nu)
\end{array}
\right)
\left(
\begin{array}{cc}
\cos^2\theta_m & \sin^2\theta_m \\
\sin^2\theta_m & \cos^2\theta_m 
\end{array}
\right)
\left(
\begin{array}{c}
1\\
0
\end{array}
\right)\ ,
\label{KuPa2}
\end{equation}
%...........................................................................
where $P_c(\nu)$ is the so-called crossing probability for neutrinos%
%-------------------------------------------
\footnote{The cases $P_c\simeq0$ and $P_c\neq 0$ discriminate  adiabatic and
nonadiabatic transitions in matter.}
%-------------------------------------------
 $[P_c(\nu)=P(\nu_{2m}\to\nu_1)]$, and $\theta_m$ is the effective mixing
angle in matter at the origin, defined by 
%...........................................................................
\begin{eqnarray}
\sin2\theta_m &=& \frac{\sin2\theta}
{\sqrt{(\cos2\theta-V/k)^2+\sin^2 2\theta}} \label{s2m}\ ,\\
\cos2\theta_m &=& \frac{\cos2\theta-V/k}
{\sqrt{(\cos2\theta-V/k)^2+\sin^2 2\theta}} \label{c2m}\ .
\end{eqnarray}
%...........................................................................
Note that in Eq.~(\ref{KuPa2}) it is understood that oscillating terms are
averaged out, thus providing an incoherent $\nu$ state at detection.

The high supernova core density (at the start of  neutrino free streaming)
implies $V/k\gg 1$ in Eqs.~(\ref{s2m}) and (\ref{c2m}), so that 
$\sin2\theta_m\simeq 0$ and $\cos2\theta_m\simeq -1$. From Eq.~(\ref{KuPa2}),
one can then reduce the calculation of $P_{ee}$ to that of $P_c$,
%...........................................................................
\begin{eqnarray}
P_{ee}(\nu) &=& \cos^2\theta\, P_c(\nu) + \sin^2\theta\, [1-P_c(\nu)]
\label{PeePc}\\
            &=& U^2_{e1}\, P_c(\nu) + U^2_{e2}\, [1-P_c(\nu)]\ .
\label{UPc}
\end{eqnarray}
%...........................................................................

%-----------------------------------
\subsubsection{Numerical approach}

One possible approach to the calculation of $P_c$ (and $P_{ee}$) is the
numerical integration of the neutrino evolution equations along the supernova
potential profile, as advocated in some recent works \cite{Ta00,Ta01,Sa01}, as
well as in a few earlier ones  (see, e.g., \cite{Mi88}). For the purposes of
our work, we  have performed a numerical (Runge-Kutta) calculation for $P_c$,
assuming the two potential profiles in Fig.~1.

Figures~2 and 3 show  our numerical results as dotted isolines for $P_c(\nu)$
in the mass-mixing plane $(\Delta m^2/E,\,\tan^2\theta)$, for the case of
power-law and realistic potential, respectively.%
%------------------
\footnote{The solid curves in Figs.~2 and 3 correspond to our analytical
approximation for $P_c$, as discussed later.}
%------------------
The ``bumpy'' structure of the realistic $V(x)$ profile is reflected by the
``wiggling'' behavior of $P_c$ in Fig.~3, leading to significant differences
with the $P_c$ isolines of Fig.~2 in part of the parameter space. However, one
can note that, for $\Delta m^2/E\to 0$, the detailed structure of $V(x)$ is
irrelevant, and $P_c(\nu)\to\cos^2\theta$ in both Figs.~2 and 3 (extremely
nonadiabatic limit).

Although brute-force numerical calculations of $P_c$ can provide, in
principle, ``exact'' results, it should be stressed that  they do not
represent an optimal and efficient approach in the case of supernovae.
Computer routines for integration are tipically time-consuming, being required
to track a large number of oscillation cycles along a potential profile
spanning many orders of magnitude. Instabilities and inaccuracies in the
numerical results can easily emerge for realistic potentials  (as experienced
by ourselves), as a consequence of  sudden variations in the lowest order
$V(x)$ derivatives.%
%--------------------------------
\footnote{Indeed, we think that some numerical artifacts (fake wiggles)  might
be present in the numerical calculations of $P_{ee}$ as graphically reported
in \protect\cite{Mi88}.}
%---------------------------------
Moreover, numerical integration produces additional but useless (unobservable)
information on oscillating factors and phases, and is thus inefficient for
practical purposes.%
%----------------------------------------
\footnote{For instance, the authors of \protect\cite{Ta00,Ta01} need to
time-average their numerical probabilities, in order to force an incoherent
initial state for the implementation of Earth matter effects.}
%----------------------------------------
Last, but not least, the uncertainties affecting simulated supernova density
profiles make it preferable to perform several   approximated (but quick)
calculations of $P_c$ for different trial functions $V(x)$,  rather than a
single (but time-consuming) exact numerical calculation for a fixed $V(x)$.

%------------------------------------------
\subsubsection{Analytical approach}

The previous discussion indicates that a handy analytical approximation to the
numerical results for $P_c$, applicable in the whole parameter space, is
highly desirable. We propose (and motivate below) the following analytical
recipe in three steps:\\
$(i)$ Identify the point $x_p$
where {\em the potential equals the wavenumber},
%...........................................................................
\begin{equation}
V(x_p) = k \ ;
\label{Vk}
\end{equation}
%...........................................................................
$(ii)$ calculate the so-called {\em density scale factor\/} $r$ at that point,
%...........................................................................
\begin{equation}
r = -\left[\frac{1}{V(x)}\,\frac{dV(x)}{dx}\right]^{-1}_{x=x_p}\ ;
\label{r}
\end{equation}
%...........................................................................
$(iii)$ Insert the above $r$ in the {\em double-exponential\/}
parametrization for $P_c(\nu)$ (originally derived in the context of solar
neutrinos \cite{Pe88,Kr88}),
%...........................................................................
\begin{equation}
P_c(\nu) = \frac{\exp(2\pi r k \cos^2\theta)-1}{\exp(2\pi r k)-1} \ .
\label{Pc}
\end{equation}
%...........................................................................

The results of such an exceedingly simple analytical recipe are shown as solid
isolines for $P_c$ in Fig.~2 (power-law case) and Fig.~3 (realistic case), to
be compared with the corresponding dashed isolines (exact numerical results).
The agreement between numerical and analytical estimates of $P_c$ is extremely
good in the whole parameter space, the difference being $\delta P_c = 2\times
10^{-2}$ in the worst cases.%
%---------------------------
\footnote{We have investigated a variety of supernova density profiles
available in the published or unpublished literature, and obtained similarly
good results (not shown).}
%---------------------------
The final accuracy for $P_{ee}$ is even better, since Eq.~(\ref{PeePc})
implies $\delta P_{ee}=|\cos 2\theta|\delta P_c<\delta P_c$.

Notice that, in the exact power-law case [Eq.~(\ref{pow})], the calculation of
$r$ through Eqs.~(\ref{Vk}) and (\ref{r}) is trivial,
%...........................................................................
\begin{equation}
\frac{r}{R_\odot}=\frac{1}{n}\left(\frac{V_0}{k}\right)^{1/n}\ .
\label{rpow}
\end{equation}
%...........................................................................
In the case of realistic potential profile,  the only modest complication is
the numerical solution of Eq.~(\ref{Vk}) and the evaluation of the derivative 
in Eq.~(\ref{r}) for the given (tabulated or parametrized) function $V(x)$.

Our effective analytical prescription for $P_c(\nu)$ in supernovae
[Eqs.~(\ref{Vk})--(\ref{Pc})] stems  from several recent improvements in the
understanding of nonadiabatic transitions, as discussed below. Although such
improvements have been mainly tested in solar neutrino oscillations, they are
often applicable also to supernova $\nu$ oscillations.

%---------------------------------------------------------
\subsubsection{Discussion of the analytical prescription}

Equation~(\ref{Pc}), originally derived for the solar neutrino  (exponential)
density profile \cite{Pe88,Kr88} within the unnecessary restriction 
$\theta<\pi/4$, was explicitly shown in \cite{Go01,Fr00} to hold for
$\theta\geq \pi/4$ as well, especially for appropriately chosen density scale
factors $r$ \cite{Li01}. The double-exponential form of $P_c$ for $\theta\geq
\pi/4$ has also been recently applied to the transitions of high-energy
neutrinos from the decay of hypothetical massive particles trapped in the Sun
\cite{Go00}, and to the transitions of supernova neutrinos
\cite{Ka00,Ka01,St01}. Such  parametrization for $P_c$ has thus several
desirable properties: $(i)$ It is a good ansatz (as originally advocated in
\cite{Ku89}) for generic density profiles; $(ii)$ It holds in both octants of
$\theta$; $(iii)$ It reproduces the extremely nonadiabatic limit at small $k$;
and $(iv)$ It reproduces the single-exponential,  Landau-Zener (LZ) limit at
small $\theta$ (see, e.g., \cite{KuPa}). As a final remark on Eq.~(\ref{Pc}),
it should be noted that the single- and double-exponential forms for $P_c$
involve, in general, a  function $F(\theta)$ depending on the potential
profile (see Table~I in \cite{Ku89}). Our choice in Eq.~(\ref{Pc}) corresponds
to take $F(\theta)=1-\tan^2\theta$, which is the exact result for a solar-like
(exponential) profile, and represents the leading prefactor of $F$ in a $1/n$
expansion%
%-------------------------------------------
\footnote{The exponential profile case can be seen as the $n\to \infty$ limit
of the power-law profile case \protect\cite{Ku89}.}
%------------------------------------------- 
for a supernova-like (power-law) profile \cite{Ku89}. We have checked that the
inclusion of the full (more complicated) expression for $F(\theta)$ in the
power-law case \cite{Ku89,KuPa,Ka01} does not lead to a significant
improvement in the (already high) accuracy of $P_c$ of our analytical
prescription, and thus  we advocate the simpler form for $P_c$ given in
Eq.~(\ref{Pc}) also for supernovae.

The concept of a ``running'' density scale factor $r=r(x_p)$ [as in
Eq.~(\ref{r})] was also originally introduced in the context of solar
neutrinos \cite{Kr88}, typically by calculating $r$ at the point $x_p=x_{\rm
res}$ defined by the so-called ``resonance'' condition $V(x_{\rm res})=\Delta
m^2\cos 2\theta$ (see, e.g., \cite{Ku89}). Such a choice for $x_p$,  although
successful at relatively small $\theta$, is clearly not applicable for
$\theta\geq \pi/4$ \cite{Ours},  and fails to describe correctly nonadiabatic
transitions at small $k$, where $P_c\neq 0$ at $\theta\sim\pi/4$
\cite{Fr01,Li01}. For large $\theta$, the resonance condition can be
misleading, if not meaningless, and it is more appropriate to  characterize
$P_c$ through the point $x_{\rm MVA}$ where maximum violation of adiabaticity
(MVA) is attained \cite{Fr01,Li01,Ka01}. Indeed, in the context of solar
neutrinos, the prescription $x_p=x_{\rm MVA}$ for $r(x_p)$ is more accurate
and physically more consistent than $x_p=x_{\rm res}$ \cite{Li01}. In the
context of  supernova neutrino oscillations,  it has also been recently
suggested that $x_{\rm MVA}$ might play an important role as well
\cite{Ka01,St01}, although the authors of \cite{Ka01,St01} {\em do not\/}  use
the prescription $r=r(x_{\rm MVA})$, but make an improved WKB calculation of
the LZ exponent for $P_c$ (involving a numerical integration in the complex
plane). However, we have verified that  the prescription $r=r(x_{\rm MVA})$
gives very accurate results for $P_c$ in the whole mass-mixing plane for
supernovae (very close to the solid isolines in Figs.~2 and 3), making it
unnecessary, in practice, to resort to WKB-inspired or other relatively
complicated approaches. In fact, our Eq.~(\ref{Vk}) is just a suitable
approximation of the MVA condition, as we now discuss.

For a monotonic $V(x)$ profile, the MVA point is uniquely defined in terms of
the flex point of $\cos2\theta_m(x)$ \cite{Li01},
%...........................................................................
\begin{equation}
\left( \frac{d^2\cos2\theta_m(x)}{dx^2}\right)_{x=x_{\rm MVA}} = 0\ .
\end{equation}
%...........................................................................
For a power-law profile as in Eq.~(\ref{pow}), the above equation implies that
%...........................................................................
\begin{equation}
V(x_{\rm MVA})=k\cdot g(n,\theta)\ ,
\label{Vkg}
\end{equation}
%...........................................................................
where 
%...........................................................................
\begin{equation}
g(n,\theta) = \frac{\cos 2\theta}{2}\,\frac{2-n}{1-2n}
\pm \sqrt{1-\frac{2-n}{1-2n}+
\left(\frac{\cos 2\theta}{2}\,\frac{2-n}{1-2n}\right)^2}\ ,
\end{equation}
%...........................................................................
and the $\pm$ sign must be chosen so that $g>0$. For $n$ close to $3$
(supernova case), by keeping only the leading term in the  expansion of the
square root in $g(n,\theta)$, one obtains:
%...........................................................................
\begin{eqnarray}
n=2 &\longrightarrow & g=1 \ ,\\
n=3 &\longrightarrow & g\simeq 1+\frac{\cos2\theta-1}{10} \ ,\\
n=4 &\longrightarrow& g\simeq 1 +\frac{\cos2\theta-1}{7}\ ,
\end{eqnarray}
%...........................................................................
namely, $g\simeq 1$ for $n\simeq 3\pm 1$, up to $\sim 10\%$ errors.

A fractional error $\delta$ in the evaluation of $V(x_{\rm MVA})$ leads, for
power-law potentials, to a fractional error $\delta/n$ in the evaluation of
$r(x_{\rm MVA})$. Therefore, by setting in any case $g=1$ [and thus $V=k$, as
in Eq.~(\ref{Vk}), rather than $V=kg$, as in Eq.~(\ref{Vkg})] we expect a mere
few~\% variation in the running value of $r$, which is of little relevance
when $r$ is inserted in $P_c$ through Eq.~(\ref{Pc}), as we have numerically
checked.

In conclusion, the condition $V(x_p)=k$ in Eq.~(\ref{Vk})  represents a good
approximation to the MVA condition for supernova neutrino oscillations, and
can be used to replace the time-honored (but inapplicable at large $\theta$)
resonance prescription.%
%----------------------------------
\footnote{It is curious to note that, in the context of solar neutrinos, the
relevance of the point where $V=k$ (as opposed to the resonance point)  was
suggested in \protect\cite{Pant} and then abandoned in the literature. See
also \protect\cite{Fr01} for an updated discussion.}
%----------------------------------
One has then to calculate [through Eq.~(\ref{r})] the corresponding (running)
value of the density scale factor $r=r(x_p)$ [provided by Eq.~(\ref{rpow}) for
an exact power-law profile], insert $r$ in the double-exponential form for
$P_c$ [Eq.~(\ref{Pc})], and finally get $P_{ee}$ through Eq.~(\ref{PeePc}).
Our analytical prescription is applicable to generic (power-law or realistic)
$V(x)$ profiles in the whole $2\nu$ parameter space, with a typical percent
accuracy in $P_{ee}$.

Figure~4 explicitly reports the results of such a prescription for $P_{ee}$
for the two profiles in Fig.~1, and for a representative supernova $\nu$
energy ($E=15$ MeV). Notice that the $P_{ee}$ isolines for the power-law
profile (dotted curves)  are simply calculated  through elementary functions
[Eqs.~(\ref{Vk})--(\ref{rpow})]. Notice also that the $P_{ee}$ isolines for
the realistic and power-law cases in Fig.~4 appear to differ significantly in
two regions: $(i)$ at relatively small mixing $(\tan^2\theta\lesssim 0.1)$;
and $(ii)$ at nearly maximal mixing  $(\tan^2\theta\lesssim 1)$ with $\Delta
m^2\sim 10^{-8\pm 1}$~eV$^2$. In connection with  solar neutrinos, this fact
implies that the detailed supernova density profile can be relevant for the
oscillation parameters corresponding to the so-called small mixing angle
(SMA), low $\Delta m^2$ (LOW), and quasivacuum  oscillation (QVO) solutions to
the solar neutrino problem \cite{SNO1}.  Conversely, for oscillation
parameters within the so-called large mixing angle (LMA) solution at $\Delta
m^2\gtrsim 10^{-5}$~eV$^2$, or within the multiple vacuum oscillation (VO)
solutions at  $\Delta m^2\lesssim 10^{-9}$~eV$^2$ \cite{SNO1}, transitions in
supernova matter become, respectively, purely adiabatic or extremely
nonadiabatic, with no significant dependence on the details of the $V(x)$
profile.

A final technical remark is in order. For solar neutrinos, it was shown in
\cite{Li01} that the MVA-inspired recipe for $P_c$  [$r=r(x_{\rm MVA})$]  has
to be matched and replaced, at small $k$, with $r={\em const}$
\cite{Fr00,Fr01}. The constant (limiting) value for $r$ can be elegantly
derived, through a perturbative approach \cite{Li01}, in terms of an integral
involving $V(x)$ in the convective zone of the Sun, where $V(x)$ experiences a
sudden drop. Conversely, in supernovae, $V(x)$ vanishes in a smoother way at
large $x$,  and the small-$k$ correction ($r={\em const}$) becomes unnecessary
in practice. We have numerically checked that such correction  does not
appreciably improve the  accuracy of the simple MVA-inspired recipe in
Eqs.~(\ref{Vk})--(\ref{Pc}). Therefore, concerning the calculation of $P_c$ in
supernova neutrino transitions, we advocate the use of
Eqs.~(\ref{Vk})--(\ref{Pc})  in the whole $2\nu$ parameter space.

%%%%%%%%%%%%%%%%%%%%%%%%%%%%%%%%%%%%%%%%%%%%%%%%%%%%%%%%%%%%%%%%%%%%%%%%%%%%
\subsection{$2\nu$ transitions: Antineutrinos}

The extension of our analytical prescription from the neutrino case to the
antineutrino  case can be obtained through the replacement $V/k\to -V/k$ in
$P_{ee}$. By conventionally keeping $V>0$ (and $\theta$ unaltered), this
implies
%...........................................................................
\begin{equation}
P_{ee}(\overline\nu|+\Delta m^2)\equiv P_{ee}(\nu|-\Delta m^2)\ .
\end{equation}
%...........................................................................
The change of sign of $\Delta m^2$ for neutrinos is equivalent to a swap of the
mass labels $1\leftrightarrow 2$,
%...........................................................................
\begin{equation}
P_{ee}(\nu|-\Delta m^2)\equiv P_{ee}(\nu|+\Delta m^2)_{1\leftrightarrow2}\ ,
\end{equation}
%...........................................................................
corresponding to $U^2_{e1}\leftrightarrow U^2_{e2}$ and
$\sin^2\theta\leftrightarrow \cos^2\theta$.%
%-------------------------------
\footnote{Such well-known $2\nu$ symmetry properties are repeated here, in
preparation of the more complicated $3\nu$ case.}
%------------------------------

The above two equations imply that, for fixed $\Delta m^2>0$
[Eq.~(\ref{Deltam})], $P_{ee}(\overline\nu)$ can be obtained  from
Eqs.~(\ref{PeePc}) and (\ref{Pc})  through a $1\leftrightarrow 2$ swap,
namely,
%...........................................................................
\begin{eqnarray}
P_{ee}(\overline\nu) &\equiv& P_{ee}(\nu)_{1\leftrightarrow 2}\nonumber\\
&=& \sin^2\theta\, P_c(\overline\nu)+\cos^2\theta\, [1-P_c(\overline\nu)]\ ,
\end{eqnarray}
%...........................................................................
where
%...........................................................................
\begin{eqnarray}
P_c(\overline \nu) &\equiv& P_c(\nu)_{1\leftrightarrow 2}\nonumber \\
&=&\frac{\exp(2\pi r k \sin^2\theta)-1}{\exp(2\pi r k)-1}  
\label{PcA}\ ,
\end{eqnarray}
%...........................................................................
with $r$ [Eq.~(\ref{r})] to be evaluated at the same ($\theta$-independent)
point $x_p$ defined for the neutrino case in Eq.~(\ref{Vk}).  Isolines of
$P_c(\overline\nu)$ from Eq.~(\ref{PcA}) are just the mirror images (around
the axis $\tan^2\theta=1$) of those obtained for $P_c(\nu)$ in Figs.~2 and 3.

%%%%%%%%%%%%%%%%%%%%%%%%%%%%%%%%%%%%%%%%%%%%%%%%%%%%%%%%%%%%%%%%%%%%%%%%%%%%
\subsection{$2\nu$ transitions: Summary}

A neat summary of the previous analytical results for $\nu$ and $\overline
\nu$ can be obtained by introducing a new notation,
%...........................................................................
\begin{eqnarray}
P_c^\pm &\equiv & \frac{\exp(\pm 2\pi r k \cos^2\theta)-1}
{\exp(\pm2\pi r k)-1} \\
&=&\left\{\begin{array}{cc}
P_c(\nu)\  & (+)\ ,\\
1-P_c(\overline \nu)\ & (-)\ .
\end{array}
\right. 
\end{eqnarray}
%...........................................................................
In terms of $P_c^{\pm}$, the expressions of $P_{ee}(\nu)$ and
$P_{ee}(\overline \nu)$ for $2\nu$ transitions are unified as 
%...........................................................................
\begin{equation}
P_{ee}^{2\nu} = U^2_{e1}\, P_c^\pm + U^2_{e2}\,(1-P_c^\pm)\ ,
\label{Pee2nu}
\end{equation}
%...........................................................................
where the $+$ sign applies to neutrinos, while the $-$ sign to antineutrinos.

In the above equation, for  any sign $(\pm)$, both $V$ and $k$ are kept $>0$.
All physical cases are covered by letting $\theta$ span its whole range
$[0,\,\pi/2]$. The density scale factor $r$ [Eq.~(\ref{r})] is calculated at
the point $x_p$ where $V(x_p)=k$ [Eq.~(\ref{Vk})].  The accuracy of such
analytical prescription, as compared with exact numerical results, is at the
percent level for both realistic and power-law density profiles, as
demonstrated in Figs.~2 and 3 for the representative cases shown in Fig.~1. In
the power-law case [Eq.~(\ref{pow})], the calculation of $r$ is further
simplified [Eq.~(\ref{rpow})].

%%%%%%%%%%%%%%%%%%%%%%%%%%%%%%%%%%%%%%%%%%%%%%%%%%%%%%%%%%%%%%%%%%%%%%%%%%%%
\section{Three-flavor transitions with hierarchical mass spectra}
%%%%%%%%%%%%%%%%%%%%%%%%%%%%%%%%%%%%%%%%%%%%%%%%%%%%%%%%%%%%%%%%%%%%%%%%%%%%

Supernova neutrinos can provide peculiar tests of three-flavor oscillations in
matter, owing to the wide dynamical range of $V(x)$ in collapsing stars  (see
\cite{Di00,Lu00,Lu01,Mi01,Du01,Du02,Du03,Ku00,Ta00,Ta01,Sa01} and references
therein for recent $3\nu$ studies).  It is thus important to generalize the
previous $2\nu$ results to the case of $3\nu$ transitions, as we do in this
section for the phenomenologically interesting cases with hierarchical mass
spectra.  We think it useful to review the derivation of $P_{ee}^{3\nu}$
(recovering some known results) with no reference to the often (mis)used
concept of ``resonant transition.''

%%%%%%%%%%%%%%%%%%%%%%%%%%%%%%%%%%%%%%%%%%%%%%%%%%%%%%%%%%%%%%%%%%%%%%%%%%%%
\subsection{$3\nu$ transitions: Notation and phenomenological input}

We assume mixing among three active neutrinos $(\nu_e,\,\nu_\mu,\,\nu_\tau)$
and three mass eigenstates $(\nu_1,\,\nu_2,\,\nu_3)$ through a unitary 
matrix%
%---------------
\footnote{In the context of supernova neutrinos, $U$ can be taken real without
loss of generality.}
%----------------
 $U$. The matrix elements $U_{ei}$ relevant for $P_{ee}^{3\nu}$ are
parametrized in terms of two mixing angles
$(\phi,\,\omega)=(\theta_{13},\,\theta_{12})$ \cite{KuPa},
%...........................................................................
\begin{eqnarray}
U^2_{e1} &=& \cos^2\phi\,\cos^2\omega\ ,\label{Ue1}\\
U^2_{e2} &=& \cos^2\phi\,\sin^2\omega\ ,\label{Ue2}\\
U^2_{e3} &=& \sin^2\phi\ .\label{Ue3}
\end{eqnarray}
%...........................................................................

The kinematical parameters are completed by two independent squared mass
differences $(\delta m^2,\,m^2)$. The dynamics is fixed by $V(x)$, and the
full supernova $3\nu$ parameter space $S_{3\nu}$ is 
%...........................................................................
\begin{equation}
S_{3\nu}=(\delta m^2,\,m^2,\,\omega,\,\phi,\,V) \ .\label{S3nu}
\end{equation}
%...........................................................................

Solar and reactor neutrino oscillation analyses suggest  \cite{SNO1,Da01} 
%...........................................................................
\begin{equation}
\delta m^2=|m^2_2-m^2_1|\lesssim 7\times 10^{-4} {\rm\ eV}^2\ ,
\label{sol}
\end{equation}
%...........................................................................
while atmospheric neutrino analyses indicate \cite{Da01,Subd} 
%...........................................................................
\begin{equation}
m^2\simeq |m^2_3-m^2_{1,2}|\sim 3\times 10^{-3} {\rm\ eV}^2\ ,
\label{atm}
\end{equation}
%...........................................................................
thus favoring the so-called hierarchical hypothesis \cite{Comp}
%...........................................................................
\begin{equation}
\delta m^2 \ll m^2\ ,
\label{hier}
\end{equation}
%...........................................................................
very often used in the literature.

Under the assumption of Eq.~(\ref{hier}), we parametrize the mass spectra (up
to an overall mass scale) as
%...........................................................................
\begin{equation}
(m^2_1,m^2_2,m^2_3)   = \left\{
\begin{array}{ccc}
\left( -\frac{\delta m^2}{2},\,+\frac{\delta m^2}{2},\,+m^2\right)
&,&{\rm direct\ hierarchy}\ ,\\[4mm]
\left( -\frac{\delta m^2}{2},\,+\frac{\delta m^2}{2},\,-m^2\right)
&,&{\rm inverse\ hierarchy}\ ,
\end{array}
\right.
\label{spec}
\end{equation}
%...........................................................................
where, conventionally, $m^2_3-m^2_{1,2}>0$ ($<0$) identifies the so-called
case of direct (inverse) hierarchy,  while $m^2_2-m^2_1>0$ for both
hierarchies. As far as $\phi,\omega\in[0,\,\pi/2]$, such a convention can be
shown to cover all physical cases, both in vacuum and in matter \cite{Gl01}.

Besides Eqs.~(\ref{sol})--(\ref{hier}), a further phenomenological input comes
from the combination of reactor and atmospheric neutrino data, providing
\cite{Da01,Subd}
%...........................................................................
\begin{equation}
\sin^2\phi=U^2_{e3}\lesssim {\rm few\ }\%\ .
\label{CHOOZ}
\end{equation}
%...........................................................................

As a consequence of the hierarchical assumption of Eq.~(\ref{hier}) [and, to
some extent, also of Eq.~(\ref{CHOOZ})], the $3\nu$ dynamics approximately
reduces to the dynamics in two $2\nu$ subsystems, dominated by relatively low
$(L)$ and high $(H)$ values of the matter density, according to parameter
space factorization
%...........................................................................
\begin{equation}
S_{3\nu}\simeq L_{2\nu}\otimes H_{2\nu}\ = 
(\delta m^2,\,\omega,\,V\cos^2\phi)\otimes
(m^2,\,\phi,\,V)
\label{fact}
\end{equation}
%...........................................................................
(see \cite{KuPa} and references therein). The corresponding ``low'' and
``high'' neutrino oscillation wavenumbers are defined by
%...........................................................................
\begin{equation}
k_L={\delta m^2}/2E
\label{kL}
\end{equation}
%...........................................................................
and
%...........................................................................
\begin{equation}
k_H=m^2/2E\ ,
\label{}
\end{equation}
%...........................................................................
respectively.

%%%%%%%%%%%%%%%%%%%%%%%%%%%%%%%%%%%%%%%%%%%%%%%%%%%%%%%%%%%%%%%%%%%%%%%%%%%%
\subsection{$3\nu$ transitions: Neutrinos with direct hierarchy}

For neutrinos with direct mass hierarchy, the factorization of dynamics in
Eq.~(\ref{fact}) leads to \cite{KuPa}
%...........................................................................
\begin{eqnarray}
P_{ee}(\nu) &=&
\left(
\begin{array}{ccc}
U^2_{e1}\,,& U^2_{e2}\,,& U^2_{e3}
\end{array}
\right)
\left(
\begin{array}{ccc}
1-P_L(\nu)& P_L(\nu)& 0\\
P_L(\nu)& 1-P_L(\nu)&0 \\
0& 0& 1 
\end{array}
\right)\nonumber\\
&&\times
\left(
\begin{array}{ccc}
1&0 &0 \\
0&1-P_H(\nu) &P_H(\nu) \\
0& P_H(\nu)&1-P_H(\nu) 
\end{array}
\right)
\left(
\begin{array}{c}
U^2_{e1,m} \\
U^2_{e2,m} \\
U^2_{e3,m}
\end{array}
\right)\ ,
\label{Pfact}
\end{eqnarray}
%...........................................................................
where $P_L$ and $P_H$ are the crossing probabilities for the low and high
density transitions, respectively, and the elements $U^2_{ei,m}$ in matter are
defined in analogy to Eqs.~(\ref{Ue1})--(\ref{Ue3}), but with  neutrino mixing
angles $\omega_m$ and $\phi_m$ in matter given by
%...........................................................................
\begin{eqnarray}
\sin2\omega_m &=& \frac{\sin2\omega}
{\sqrt{(\cos2\omega-\cos^2\phi\,V/k_L)^2+\sin^2 2\omega}}\ ,\label{swm}\\
\cos2\omega_m &=& \frac{\cos2\omega-V/k_L}
{\sqrt{(\cos2\omega-\cos^2\phi\,V/k_L)^2+\sin^2 2\omega}}\ ,\label{cwm}
\end{eqnarray}
%...........................................................................
and
%...........................................................................
\begin{eqnarray}
\sin2\phi_m &=& \frac{\sin2\phi}
{\sqrt{(\cos2\omega-V/k_H)^2+\sin^2 2\phi}}\ ,\label{sfm}\\
\cos2\phi_m &=& \frac{\cos2\phi-V/k_H}
{\sqrt{(\cos2\omega-V/k_H)^2+\sin^2 2\phi}}\ ,\label{cfm}\ 
\end{eqnarray}
%...........................................................................
at zeroth order in $\delta m^2/m^2$ \cite{KuPa}. The (high density) initial
condition $V/k_{L,H}\gg 1$ leads then to $\cos2\phi_m\simeq -1\simeq
\cos2\omega_m$ and to the known expression
%...........................................................................
\begin{equation}
P_{ee}(\nu)=U^2_{e1}\,P_L(\nu)\,P_H(\nu)+U^2_{e2}\,[1-P_L(\nu)]\,P_H(\nu)+
U^2_{e3}\,[1-P_H(\nu)]\ .
\label{Pnudir}
\end{equation}
%...........................................................................

On the basis of Eq.~(\ref{fact}) and of our $2\nu$ prescription in 
Eqs.~(\ref{Vk})--(\ref{Pc}), $P_L(\nu)$ and $P_H(\nu)$ can be analytically
expressed as
%...........................................................................
\begin{equation}
P_L(\nu)=\frac{\exp(2\pi r_Lk_L\cos^2\omega)-1}{\exp(2\pi r_Lk_L)-1}\ 
\label{PLnu}
\end{equation}
%...........................................................................
and
%...........................................................................
\begin{equation}
P_H(\nu)=\frac{\exp(2\pi r_Hk_H\cos^2\phi)-1}{\exp(2\pi r_Hk_H)-1}\ ,
\label{PHnu}
\end{equation}
%...........................................................................
where 
%...........................................................................
\begin{equation}
r_{L,H}=-\left[\frac{1}{V(x)}\,\frac{dV(x)}{dx}\right]^{-1}_{x=x_{L,H}}\ ,
\label{rHL}
\end{equation}
%...........................................................................
and the points $x_{L}$ and $x_H$ are defined by the $V=k$ condition%
%------------------
\footnote{In principle, Eq.~(\protect\ref{fact}) implies an effective
potential $V(x)\cos^2\phi$ for the $L$ subsystem, and thus  $V(x_L)\cos^2\phi
= k_L$. However, we have checked that, within the phenomenological bound in
Eq.~(\protect\ref{CHOOZ}), the resulting difference in the calculation of
$P_{ee}$ is completely negligible. We prefer then the simpler condition
$V(x_L)=k_L$, analogous to $V(x_H)=k_H$ for the $H$ subsystem.}
%------------------
 [Eq.~(\ref{Vk})]
%...........................................................................
\begin{equation}
V(x_{L,H})=k_{L,H}\ .
\label{VkHL}
\end{equation}
%...........................................................................
Equations (\ref{Pnudir})--(\ref{VkHL}) allow the calculation of $P_{ee}(\nu)$
in the whole $3\nu$ parameter space for direct hierarchy.

%%%%%%%%%%%%%%%%%%%%%%%%%%%%%%%%%%%%%%%%%%%%%%%%%%%%%%%%%%%%%%%%%%%%%%%%%%%%
\subsection{$3\nu$ transitions: Antineutrinos with direct  hierarchy}

The antineutrino case can be obtained in analogy to the neutrino case
[Eqs.~(\ref{Pfact})--(\ref{cfm})] through the replacement
$V/k_{L,H}\to-V/k_{L,H}$. In particular, the {\em antineutrino\/} mixing
angles in matter are given by
%...........................................................................
\begin{eqnarray}
\sin2\overline\omega_m &=& \frac{\sin2\omega}
{\sqrt{(\cos2\omega+V\cos^2\phi/k_L)^2+\sin^2 2\omega}}\ ,\label{Aswm}\\
\cos2\overline\omega_m &=& \frac{\cos2\omega+V/k_L}
{\sqrt{(\cos2\omega+V\cos^2\phi/k_L)^2+\sin^2 2\omega}}\ ,\label{Acwm}
\end{eqnarray}
%...........................................................................
and analogously for $\overline\phi_m$. The initial  high-density condition
gives  $\cos2\overline\phi_m\simeq +1\simeq \cos2\overline\omega_m$ and leads
to
%...........................................................................
\begin{equation}
P_{ee}(\overline\nu)=
U^2_{e1}\,[1-P_L(\overline\nu)]+U^2_{e2}\,P_L(\overline\nu)\ .
\label{PAnudir}
\end{equation}
%...........................................................................
Our analytical prescription for $P_L(\overline \nu)$ is then obtained, {\em
mutatis mutandis}, from  Eq.~(\ref{PcA}),
%...........................................................................
\begin{equation}
P_L(\overline\nu)=\frac{\exp(2\pi r_Lk_L\sin^2\omega)-1}{\exp(2\pi r_Lk_L)-1}
\label{PLA}\ ,
\end{equation}
%...........................................................................
with $r_L$  defined as in Eqs.~(\ref{rHL}) and (\ref{VkHL}).

%%%%%%%%%%%%%%%%%%%%%%%%%%%%%%%%%%%%%%%%%%%%%%%%%%%%%%%%%%%%%%%%%%%%%%%%%%%%
\subsection{$3\nu$ transitions: Neutrinos with inverse  hierarchy}

For fixed mass spectrum and mixing angles, neutrino and antineutrino
probabilities can be transformed one into the other by flipping the signs of
$V/k_{L,H}$ or, equivalently, by flipping the signs attached to $\delta m^2$
and $m^2$, while keeping $V>0$. The $\pm\delta m^2$ sign flip is equivalent to
the $1\leftrightarrow 2$ swap of mass labels (as discussed for the $2\nu$
case), leading to $U^2_{e1}\leftrightarrow U^2_{e2}$ [namely, 
$\sin^2\omega\leftrightarrow \cos^2\omega$, with no change in $\phi$ or
$U^2_{e3}$]. The $\pm m^2$ sign flip is instead equivalent to swap hierarchy
[see Eq.~(\ref{spec})].

From such symmetry properties one  obtains
%...........................................................................
\begin{equation}
P_{ee}(\nu\,|+\delta m^2,\,-m^2)\equiv
P_{ee}(\overline\nu|-\delta m^2,\,+m^2)\equiv
P_{ee}(\overline\nu\,|+\delta m^2,\,+m^2)_{1\leftrightarrow 2}\ ,
\label{symm1}
\end{equation}
%...........................................................................
namely, the ${\nu}_e$ survival probability for inverse hierarchy equals the
${\overline\nu}_e$ survival probability for direct hierarchy
[Eq.~(\ref{PAnudir})], under the substitution $U^2_{e1}\leftrightarrow
U^2_{e2}$,
%...........................................................................
\begin{equation}
P_{ee}(\nu)=U^2_{e2}\,[1-P_L(\nu)]+U^2_{e1}\,P_L(\nu)\ ,
\label{Pnuinv}
\end{equation}
%...........................................................................
with $P_L(\nu)$ defined as in Eq.~(\ref{PLnu}).

%%%%%%%%%%%%%%%%%%%%%%%%%%%%%%%%%%%%%%%%%%%%%%%%%%%%%%%%%%%%%%%%%%%%%%%%%%%%
\subsection{$3\nu$ transitions: Antineutrinos with inverse  hierarchy}

In analogy with the previous subsection, it can be easily realized that
%...........................................................................
\begin{equation}
P_{ee}(\overline\nu\,|+\delta m^2,\,-m^2)\equiv
P_{ee}(\nu|-\delta m^2,\,+m^2)\equiv
P_{ee}(\nu\,|+\delta m^2,\,+m^2)_{1\leftrightarrow 2}\ ,
\label{symm2}
\end{equation}
%...........................................................................
which, applied to Eq.~(\ref{Pnudir}), gives
%...........................................................................
\begin{equation}
P_{ee}(\overline\nu)=U^2_{e2}\,P_L(\overline\nu)
\,P_H(\nu)+U^2_{e1}\,[1-P_L(\overline\nu)]\,P_H(\nu)+
U^2_{e3}\,[1-P_H(\nu)]\ ,
\label{PAnuinv}
\end{equation}
%...........................................................................
with $P_L(\overline\nu)$ and $P_H(\nu)$ defined as in Eqs~(\ref{PLA})  and
(\ref{PHnu}), respectively.

Notice that we have written $P_H(\nu)$ and not $P_H(\overline\nu)$ in
Eq.~(\ref{PAnuinv}), since the $1\leftrightarrow 2$ swap makes no change in
$P_H$, defined in terms of $m^2_3-m^2_{1,2}$ and of $U^2_{e3}$. One can then
drop the argument of $P_H$ and simply write%
%----------------------------------------
\footnote{A formal distinction between $P_H(\nu)$ and $P_H(\overline\nu)$ was
kept in the notation of \protect\cite{Di00} and then dropped in
\protect\cite{Lu01}.}
%----------------------------------------
%...........................................................................
\begin{equation}
P_H(\overline\nu)=P_H(\nu)\equiv P_H\ .
\end{equation}
%...........................................................................

%%%%%%%%%%%%%%%%%%%%%%%%%%%%%%%%%%%%%%%%%%%%%%%%%%%%%%%%%%%%%%%%%%%%%%%%%%%%
\subsection{$3\nu$ transitions: Summary I}

The $3\nu$ results in Eqs.~(\ref{Pnudir}), (\ref{PAnudir}),  (\ref{Pnuinv}),
and (\ref{PAnuinv})  can be summarized as
%...........................................................................
\begin{equation}
P^{3\nu}_{ee} = U^2_{e1}X_1 +U^2_{e2}X_2 +U^2_{e3}X_3\ ,
\label{Pee3one}
\end{equation}
%...........................................................................
where the coefficients $X_i$ are given in Table~I, in terms
of $P_L(\nu)$, $P_L(\overline\nu)$, and $P_H$.

The parametrization of $P_{ee}^{3\nu}$ given in Eq.~(\ref{Pee3one}) and
Table~I agrees with the results previously obtained in \cite{Di00,Lu01}.  In
addition, however, we have provided explicit analytical approximations for
$P_L$ [Eqs.~(\ref{PLnu}) and (\ref{PLA})] and for $P_H$ [Eq.~(\ref{PHnu})],
allowing a straightforward calculation of  $P_{ee}^{3\nu}$ in the whole
mass-mixing parameter space, for generic (realistic or power-law) potential
profiles.

%%%%%%%%%%%%%%%%%%%%%%%%%%%%%%%%%%%%%%%%%%%%%%%%%%%%%%%%%%%%%%%%%%%%%%%%%%%%
\subsection{$3\nu$ transitions: Summary II}

An alternative summary for $P_{ee}^{3\nu}$  can be obtained by introducing, in
analogy with the $2\nu$ case, the notation
%...........................................................................
\begin{eqnarray}
P_L^\pm&=&\frac{\exp(\pm 2\pi r_L k_L \cos^2\omega)-1}
{\exp(\pm 2\pi r_L k_L)-1}
\label{PLpm}\\
&=&\left\{ 
\begin{array}{cc}
P_L(\nu) & (+)\ , \\
1-P_L(\overline\nu) & (-)\ ,
\end{array}
\right.
\end{eqnarray}
%...........................................................................
and
%...........................................................................
\begin{eqnarray}
P_H^\pm&=&\frac{\exp(\pm 2\pi r_H k_H \cos^2\phi)-1}{\exp(\pm 2\pi r_H k_H)-1}
\label{PHpm}\\
&\simeq&\left\{ 
\begin{array}{cc}
\exp(-2\pi r_H k_H \sin^2\phi) & (+)\ , \\
1 & (-)\ ,
\end{array}
\right.
\end{eqnarray}
%...........................................................................
where, in the last equation, we have used the phenomenological inputs in
Eqs.~(\ref{atm}) and (\ref{CHOOZ}), implying that  $2\pi r_H k_H\cos^2\phi\gg
1$ for typical supernova  potential profiles and neutrino energies.

Equation~(\ref{Pee3one}) can then be written in the equivalent form 
%...........................................................................
\begin{equation}
P_{ee}^{3\nu}=U^2_{e1}\,P_L^\pm\,P_H^\pm+U^2_{e2}\,(1-P_L^\pm)\,P_H^\pm+
U^2_{e3}\,(1-P_H^\pm)\ ,
\label{Pee3two}
\end{equation}
%...........................................................................
where the sign assignment for $P^\pm_{L,H}$ is given in Table~II for the four
possible combinations of neutrino types ($\nu$ or $\overline \nu$) and mass
hierarchy  (direct or inverse).

The neat $3\nu$ summary given in Eqs.~(\ref{PLpm}), (\ref{PHpm}),
(\ref{Pee3two})  and in Table~II makes the symmetry properties of
$P_{ee}^{3\nu}$ rather transparent. Passing from $\nu$ to $\overline\nu$, or
from direct to inverse hierarchy, appears to simply require appropriate  sign
flips.

%%%%%%%%%%%%%%%%%%%%%%%%%%%%%%%%%%%%%%%%%%%%%%%%%%%%%%%%%%%%%%%%%%%%%%%%%
\subsection{$3\nu$ transitions: Representative $P_{ee}$ calculations}

Figure~5 shows a representative analytical calculation of $P_{ee}^{3\nu}$ in
the slice $(\delta m^2,\,\tan^2\omega)$ of the $3\nu$ parameter space, which
is relevant for the $L$ transition in supernovae [Eq.~(\ref{fact})]  and for
its connection with solar neutrino oscillations. Calculations (solid lines)%
%---------------------
\footnote{The dotted curves in Fig.~5 include Earth matter effects, to be
discussed in the next Section.}
%---------------------- 
are made at fixed $E=15$~MeV, $m^2=3\times 10^{-3}$~eV$^2$, and
$\tan^2\phi=4\times 10^{-5}$ ($P_H^+=0.48$), for the power-law profile in
Fig.~1. The upper panels refer to $\nu$ (left) and $\overline\nu$ (right) in
the case of direct  hierarchy. The lower panels refer to $\nu$ (left) and
$\overline\nu$ (right) in the case of inverse hierarchy.  Notice that all
calculations for this figure involve only elementary functions,  the density
scale factor $r$ being explicitly given by Eq.~(\ref{rpow})  for a power-law
profile.

A glance at Fig.~5 shows two apparent symmetries: $P_{ee}(\nu)$ and
$P_{ee}(\overline\nu)$ for direct hierarchy  look like the mirror image of
$P_{ee}(\overline\nu)$ and $P_{ee}(\nu)$ for inverse hierarchy, respectively.
This (approximate) symmetry originates from the small value of $U^2_{e3}$ used
in the calculations of Fig.~5. Indeed, neglecting the third term (proportional
to $U^2_{e3}$) in   Eq.~(\ref{Pee3two}), and using Eq.~(\ref{Pee2nu})  (with
the identification $P_{c}^\pm=P_L^\pm$), it is
%...........................................................................
\begin{equation}
P_{ee}^{3\nu}\simeq P_H^\pm \,P_{ee}^{2\nu}\ ,
\label{mod}
\end{equation}
%...........................................................................
namely, the $3\nu$ probability is obtained from the $2\nu$ probability through
a modulation factor $P_H^\pm$. Using Eq.~(\ref{mod}) and the symmetry
properties in Eqs.~(\ref{symm1}) and (\ref{symm2}), it follows that
%...........................................................................
\begin{equation}
P_{ee}^{3\nu}(\nu,\,{\rm direct})\simeq P_H^+\,P_{ee}^{2\nu}(\nu)=
P_H^+\,P_{ee}^{2\nu}(\overline \nu)_{1\leftrightarrow 2}\simeq P_{ee}^{3\nu}
(\overline\nu,\,{\rm inverse})_{1\leftrightarrow 2}
\label{symm3}
\end{equation}
%...........................................................................
and
%...........................................................................
\begin{equation}
P_{ee}^{3\nu}(\overline\nu,\,{\rm direct})\simeq P_H^-\,P_{ee}^{2\nu}
(\overline\nu)=
P_H^-\,P_{ee}^{2\nu}(\nu)_{1\leftrightarrow 2}\simeq P_{ee}^{3\nu}
(\nu,\,{\rm inverse})_{1\leftrightarrow 2}\ ,
\label{symm4}
\end{equation}
%...........................................................................
as previously observed in Fig.~5. We remark that the above Eqs.~(\ref{symm3})
and (\ref{symm4}) represent approximate mirror symmetries, which become exact
only for $U^2_{e3}=0$. However, for $U^2_{e3}$ obeying the bound in
Eq.~(\ref{CHOOZ}), such symmetries are broken, at most, at the few percent
level.

%%%%%%%%%%%%%%%%%%%%%%%%%%%%%%%%%%%%%%%%%%%%%%%%%%%%%%%%%%%%%%%%%%%%%%%%%%%%
\section{Including Earth matter effects}
%%%%%%%%%%%%%%%%%%%%%%%%%%%%%%%%%%%%%%%%%%%%%%%%%%%%%%%%%%%%%%%%%%%%%%%%%%%%

We briefly review known analytical results about Earth matter effects, for the
sake of completeness  and self-consistency of the paper. For recent
phenomenological studies of such effects in the supernova neutrino context
see, e.g.,  \cite{Di00,Lu00,Lu01,Ta00,Ta01,Sa01}.

In general, possible Earth matter effects preceding supernova $\nu$ detection
can be implemented by the final-state substitution \cite{KuPa}
%...........................................................................
\begin{equation}
(U^2_{e1},\,U^2_{e2},\,U^2_{e3})\to (P_{e1},\,P_{e2},\,P_{e3})\ 
\label{UtoP}
\end{equation}
%...........................................................................
in Eq.~(\ref{Pee3one}) or (\ref{Pee3two}), where $P_{ei}=P(\nu_e\to\nu_i)$ in
the Earth.

Under the assumption of mass spectrum hierarchy (either direct or inverse),
the $3\nu$ calculation of $P_{ei}$ is further reduced to a $2\nu$ problem
\cite{Di00},
%...........................................................................
\begin{equation}
(P_{e1},\,P_{e2},\,P_{e3}) \simeq 
[\cos^2\phi\,(1-P_E),\,\cos^2\phi\,P_E,\,
\sin^2\phi]\ ,
\label{earth}
\end{equation}
%...........................................................................
where 
%...........................................................................
\begin{equation}
P_E = P_{e2}^{2\nu}.
\label{PE}
\end{equation}
%...........................................................................
The task is thus reduced to the $2\nu$ calculations of $P_E(\nu)$ and
$P_E(\overline\nu)$, which are independent on $\pm m^2$ and on the hierarchy
type (direct or inverse).  Analytical expressions for $P_E$  can be given for
particularly simple (or approximated) situations of Earth matter crossing.

%%%%%%%%%%%%%%%%%%%%%%%%%%%%%%%%%%%%%%%%%%%%%%%%%%%%%%%%%%%%%%%%%%%%%%%%%%%%
\subsection{One shell}

If the $\nu$ trajectory crosses only  the Earth mantle, characterized by an
approximately constant (average) density, $P_E(\nu)$ is simply given by
%...........................................................................
\begin{equation}
P_E(\nu)=\sin^2\omega + \sin2\omega_m \,\sin(2\omega_m-2\omega)\,
\sin^2\left(
\frac{k_L\,\sin2\omega}{2\,\sin2\omega_m}\,L
\right)
\end{equation}
%...........................................................................
where  $L$ is the total pathlength in the mantle, and $\omega_m$ is defined as
in Eqs.~(\ref{swm}) and (\ref{cwm}) with the appropriate potential $V$ in the
mantle [Eqs.~(\ref{V}) and (\ref{Vu})]. The antineutrino probability is
obtained through the substitution
%...........................................................................
\begin{equation}
P_E(\overline\nu)=P_E(\nu)\Big|_{V/k_L\to\,-V/k_L}\ ,
\label{replace}
\end{equation}
%...........................................................................
leading to
%...........................................................................
\begin{equation}
P_E(\overline\nu)=\sin^2\omega + \sin2\overline\omega_m \,
\sin(2\overline\omega_m-2\omega)\,
\sin^2\left(
\frac{k_L\,\sin2\omega}{2\,\sin2\overline\omega_m}\,L
\right)\ ,
\end{equation}
%...........................................................................
where ${\overline \omega}_m$ is defined through Eqs.~(\ref{Aswm}) and 
(\ref{Acwm}).

Figure~5 shows an example of  Earth matter effects on $P_{ee}^{3\nu}$ for
mantle crossing (dotted curves),  assuming $\rho =4.5$ g/cm$^3$, $Y_e=0.5$,
and $L=8500$ km (parameters which are of interest for SN1987A phenomenology;
see, e.g., \cite{Lu00,Lu01}). The neutrino potential in the mantle is then
$V_M=1.7\times 10^{-7}$ eV$^2$/MeV, implying strong effects for $V_M\sim k_L$
and thus for  $\delta m^2\sim 2EV\sim O( 10^{-6})$ eV$^2$, which may be
relevant in connection with the so-called LMA solution to the solar neutrino
problem \cite{SNO1}, as widely discussed recently
\cite{Di00,Lu00,Lu01,Ta00,Ta01,Sa01}. Notice that the approximate symmetries
in Eqs.~(\ref{symm3}) and (\ref{symm4}) are preserved by Earth matter effects,
since they act mainly on the two-neutrino $L$ transition  in hierarchical
approximation [Eq.~(\ref{earth})].

%%%%%%%%%%%%%%%%%%%%%%%%%%%%%%%%%%%%%%%%%%%%%%%%%%%%%%%%%%%%%%%%%%%%%%%%%%%%
\subsection{Multiple shells}

Neutrino oscillations across two Earth shells with different densities [mantle
$(M)$ + core $(C)$] were considered in  \cite{Mi87} in the context of
supernovae,  and intensively studied in \cite{Petc,Akhm} on general grounds,
with emphasis on interesting interference properties peculiar to layered
matter (see also \cite{Fish}). Adapting, e.g., the notation of \cite{Akhm} to
ours, the expression of  $P_E(\nu)$ for a mantle+core+mantle path in the Earth
($L=L_M+L_C+L_M$) reads 
%...........................................................................
\begin{equation}
P_E(\nu)=\sin^2\omega + W_1\, (W_1\,\cos2\omega+W_3\,\sin2\omega)\ ,
\end{equation}
%...........................................................................
where $W_{1,3}$ are the first and third component of the vector 
%...........................................................................
\begin{equation}
{\mathbf W}=2\,S_M\,Y\,{\mathbf n}_M+S_C\,{\mathbf n}_C\ ,
\end{equation}
%...........................................................................
having defined $Y$ as
%...........................................................................
\begin{equation}
Y = C_M\, C_C - ({\mathbf n}_M\cdot {\mathbf n}_C)\,S_M\,S_C\ ,
\end{equation}
%...........................................................................
the vectors ${\mathbf n}_{M}$ and  ${\mathbf n}_{C}$ as
%...........................................................................
\begin{equation}
{\mathbf n}_{M,C} = (\sin2\omega_{M,C},\,0,\,-\cos2\omega_{M,C})\ ,
\end{equation}
%...........................................................................
and $C_M$ and $C_C$ as
%...........................................................................
\begin{equation}
C_{M,C}=\cos\left(\frac{k_{L}\,\sin2\omega}{2\,\sin 2\omega_{M,C}}\,
L_{M,C}\right)
\end{equation}
%...........................................................................
(and similarly for $S_{M,C}$, with $\cos\to\sin$), where $\omega_M$ and
$\omega_C$ are the effective neutrino mixing angles in the mantle and in the
core. $P_E(\overline\nu)$ is then obtained from $P_E(\nu)$  through the
replacement indicated in Eq.~(\ref{replace}).

In the general case of $N$ different shells, not necessarily with constant
density in each shell,  the calculation of $P_E$ can also be performed
analytically, through the perturbative approach developed in \cite{Eart} in
the context of solar neutrinos.

%%%%%%%%%%%%%%%%%%%%%%%%%%%%%%%%%%%%%%%%%%%%%%%%%%%%%%%%%%%%%%%%%%%%%%%%%%%%
\section{Summary and conclusions}
%%%%%%%%%%%%%%%%%%%%%%%%%%%%%%%%%%%%%%%%%%%%%%%%%%%%%%%%%%%%%%%%%%%%%%%%%%%%

In the context of two-flavor (anti)neutrino transitions in supernovae,  we
have described a simple and accurate analytical prescription for the
calculation of the survival probability $P_{ee}$, based on a double
exponential form for the crossing probability, and inspired by the condition
of maximum violation of adiabaticity. The prescription holds in the whole
oscillation parameter space and for generic supernova density profiles. The
analytical approach has then been generalized to cover three-flavor
transitions with mass spectrum hierarchy (either direct or inverse), and to
include Earth matter effects.

The final prescription for $P_{ee}^{3\nu}$ can be summarized as follows:
%::::::::::::::::::::::::::::::::::::::::::::::::::::::::::::::::::::::
\begin{enumerate} 
\item  Assume a supernova potential profile $V(x)>0$; \item  Fix the mixing
angles $\phi=\theta_{13}$ [obeying Eq.~(\ref{CHOOZ})] and $\omega=\theta_{12}$,
and calculate  the matrix elements $U^2_{ei}$ through
Eqs.~(\ref{Ue1})--(\ref{Ue2}); 
\item  Fix  the ``solar'' and ``atmospheric''  squared mass differences,
$\delta m^2>0$ and $m^2>0$, respectively [within the phenomenological
restrictions in Eqs.~(\ref{sol})--(\ref{hier})]; 
\item At a given (anti)neutrino energy $E$,  find the points $x_L$ and $x_H$
where the potential $V$ equals the wavenumbers $k_L=\delta m^2/2E$ and
$k_H=m^2/2E$, respectively:
%..........................
$$
V(x_{L,H})=k_{L,H}\ ;
$$
%............................
%
\item Calculate the corresponding density scale factors $r_{L}$ and $r_H$ as
%......................................
$$  
r_{L,H}=-\left[\frac{1}{V(x)}\,\frac{dV(x)}{dx}\right]^{-1}_{x=x_{L,H}}
$$
%.......................................
(both $>0$ for monotonically decreasing $V$); 
\item Assign the $\pm$ signs in $P_L^\pm$ and $P_H^\pm$ by choosing the
neutrino type  ($\nu$ or $\overline \nu$) and hierarchy (direct or inverse) in
Table~II, and calculate then $P_{L,H}^\pm$ through
%...........................................................................
$$
P_L^\pm=\frac{\exp(\pm 2\pi r_L k_L \cos^2\omega)-1}
{\exp(\pm 2\pi r_L k_L)-1}
$$
%...........................................................................
and
%...........................................................................
$$
P_H^\pm=\frac{\exp(\pm 2\pi r_H k_H \cos^2\phi)-1}
{\exp(\pm 2\pi r_H k_H)-1}\ ;
$$
%...........................................................................
%
\item Calculate $P_{ee}^{3\nu}$ as
%...........................................................................
$$
P_{ee}^{3\nu}=U^2_{e1}\,P_L^\pm\,P_H^\pm+U^2_{e2}\,(1-P_L^\pm)\,P_H^\pm+
U^2_{e3}\,(1-P_H^\pm)\ ;
$$
%...........................................................................
%
\item Finally, include possible Earth matter effects as reviewed
in Sec.~IV.
\end{enumerate}
%::::::::::::::::::::::::::::::::::::::::::::::::::::::::::::
We think that such analytical  prescription may be useful to simplify the
calculation (and to help the understanding) of supernova neutrino oscillation
effects.

%%%%%%%%%%%%%%%%%%%%%%%%%%%%%%%%%%%%%%%%%%%%%%%%%%%%%%%%%%%%%%%%%%%%%%%%%%%%
\acknowledgments
%%%%%%%%%%%%%%%%%%%%%%%%%%%%%%%%%%%%%%%%%%%%%%%%%%%%%%%%%%%%%%%%%%%%%%%%%%%%

E.L.\ thanks G.\ Raffelt and H.\ Minakata for useful discussions on supernova
neutrinos during the TAUP~2001 Conference. This work was supported in part by
the Italian  {\em Istituto Nazionale di Fisica Nucleare\/} (INFN) and {\em
Ministero dell'Istruzione, dell'Universit\`a e  della Ricerca\/} (MIUR) under
the  project ``Fisica Astroparticellare.''

%%%%%%%%%%%%%% TABLE I %%%%%%%%%%%%%%%%%%%%%%%%%%%%%%%%%%%%%%%%%%%%%%%%%%%%
%%%%%%%%%%%%%%%%%%%%%%%%%%%%%%%%%%%%%%%%%%%%%%%%%%%%%%%%%%%%%%%%%%%%%%%%%%%
%...........................................................................
\begin{table}[t]
\caption{Coefficients $X_i$ to be used in the parametrization of 
$P_{ee}^{3\nu}$ given in Eq.~(\protect\ref{Pee3one}), according to the four
possible combinations of neutrino types ($\nu$ or $\overline \nu$) and mass
spectrum hierarchy (direct or inverse). The coefficients agree with those
derived in Refs.~\protect\cite{Di00,Lu01}.}
%==========================================================================
\begin{tabular}{ccccc}
Type &Hierarchy & $X_1$  & $X_2$         & $X_3$ \\
\tableline
%---------------------------------------------------------------------------
$\nu$ 		& direct & $P_L(\nu)\,P_H$ & $[1-P_L(\nu)]\,P_H$ & $1-P_H$ \\
$\overline\nu$ 	& direct & $1-P_L(\overline\nu)$ & $P_L(\overline\nu)$ &0 \\
$\nu$ 		& inverse& $P_L(\nu)$ & $1-P_L(\nu)$ & 0 \\
$\overline\nu$ 	& inverse& $[1-P_L(\overline\nu)]\,P_H$ & 
				$P_L(\overline\nu)\,P_H$ & $1-P_H$ 
\end{tabular}
%============================================================================
\end{table}
%...........................................................................

%%%%%%%%%%%%%% TABLE II %%%%%%%%%%%%%%%%%%%%%%%%%%%%%%%%%%%%%%%%%%%%%%%%%%%
%%%%%%%%%%%%%%%%%%%%%%%%%%%%%%%%%%%%%%%%%%%%%%%%%%%%%%%%%%%%%%%%%%%%%%%%%%%
%...........................................................................
\begin{table}[h]
\caption{Signs assigned to $P_L^\pm$ and $P_H^\pm$, to be used in the
parametrization of  $P_{ee}^{3\nu}$ given in Eq.~(\protect\ref{Pee3two}),
according to the four possible combinations of neutrino types ($\nu$ or
$\overline \nu$) and mass spectrum hierarchy (direct or inverse).}
%========================================================================
\begin{tabular}{cccc}
Type &Hierarchy & $P_L^\pm$  & $P_H^\pm$          \\
\tableline
%-------------------------------------------------------------------------
$\nu$ 		& direct & $+$ & $+$ \\
$\overline\nu$ 	& direct & $-$ & $-$\\
$\nu$ 		& inverse& $+$ & $-$\\
$\overline\nu$ 	& inverse& $-$ & $+$
\end{tabular}
%===========================================================================
\end{table}
%...........................................................................

%%%%%%%%%%%%%%%%%%%%% REFERENCES %%%%%%%%%%%%%%%%%%%%%%%%%%%%%%%%%%%%%%%%%%%%%

%
%%%%%%%%%%%%%%%%%%%%%%%%%%%%%%%%%%%%%%%%%%%%%%%%%%%%%%%%%%%%%%%%%%%%%%%%%%%%%%%
%%%%%%%          P O S T S C R I P T       F I G U R E S 
%%%%%%%   memo:  to include them add epsfig in the \documentstyle
%%%%%%%          and move this part before \end{document}. 
%%%%%%%          Include the following \newcommand:
%%----------------------------------------------------------------------------
\newcommand{\InsertFigure}[2]{\newpage\begin{center}\mbox{%
\epsfig{bbllx=2.2truecm,bblly=2.5truecm,bburx=20truecm,bbury=25truecm,
height=20.truecm,figure=#1}}\end{center}\vspace*{-.6truecm}%
\parbox[t]{\hsize}{\small\baselineskip=0.5truecm\hspace*{0.5truecm} #2}}
%
%----------------------------------------------------------------------------
%%%%%%%%%%%%%%%%%%%%%%%%%%%%%%%%%%%%%%%%%%%%%%%%%%%%%%%%%%%%%%%%%%%%%%%%%%%%%%%
%..............................................................................
\InsertFigure{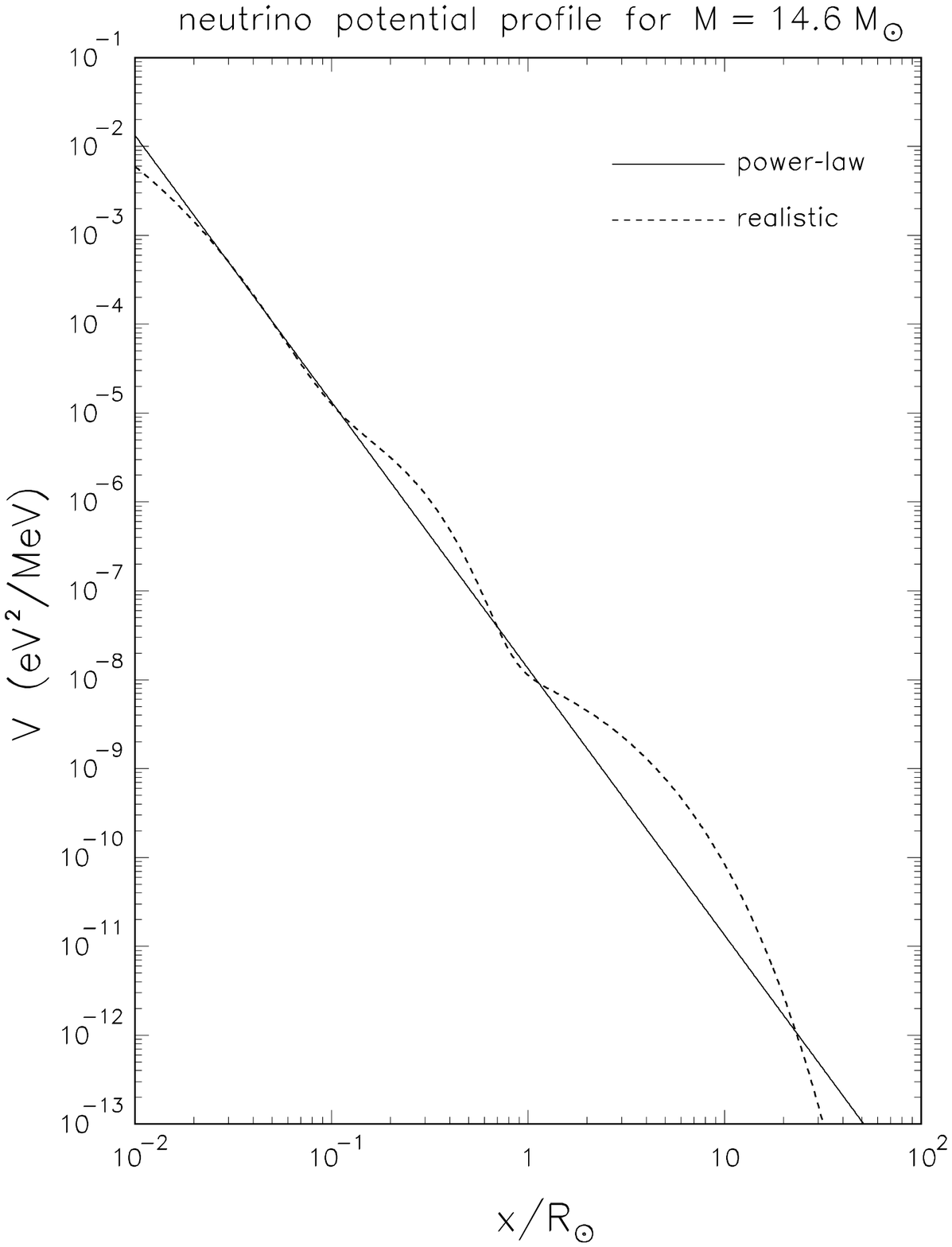}% 
{Fig.~1. Neutrino potential profiles $V(x)$ considered in this work.
Dashed curve: ``realistic'' potential, as graphically reduced from the
supernova simulation performed in \protect\cite{Sh90} assuming $14.6 M_\odot$
for the ejecta. Solid line: ``power-law'' potential $(V\propto x^{-3})$ which
best fits the realistic one.}
%..............................................................................
\InsertFigure{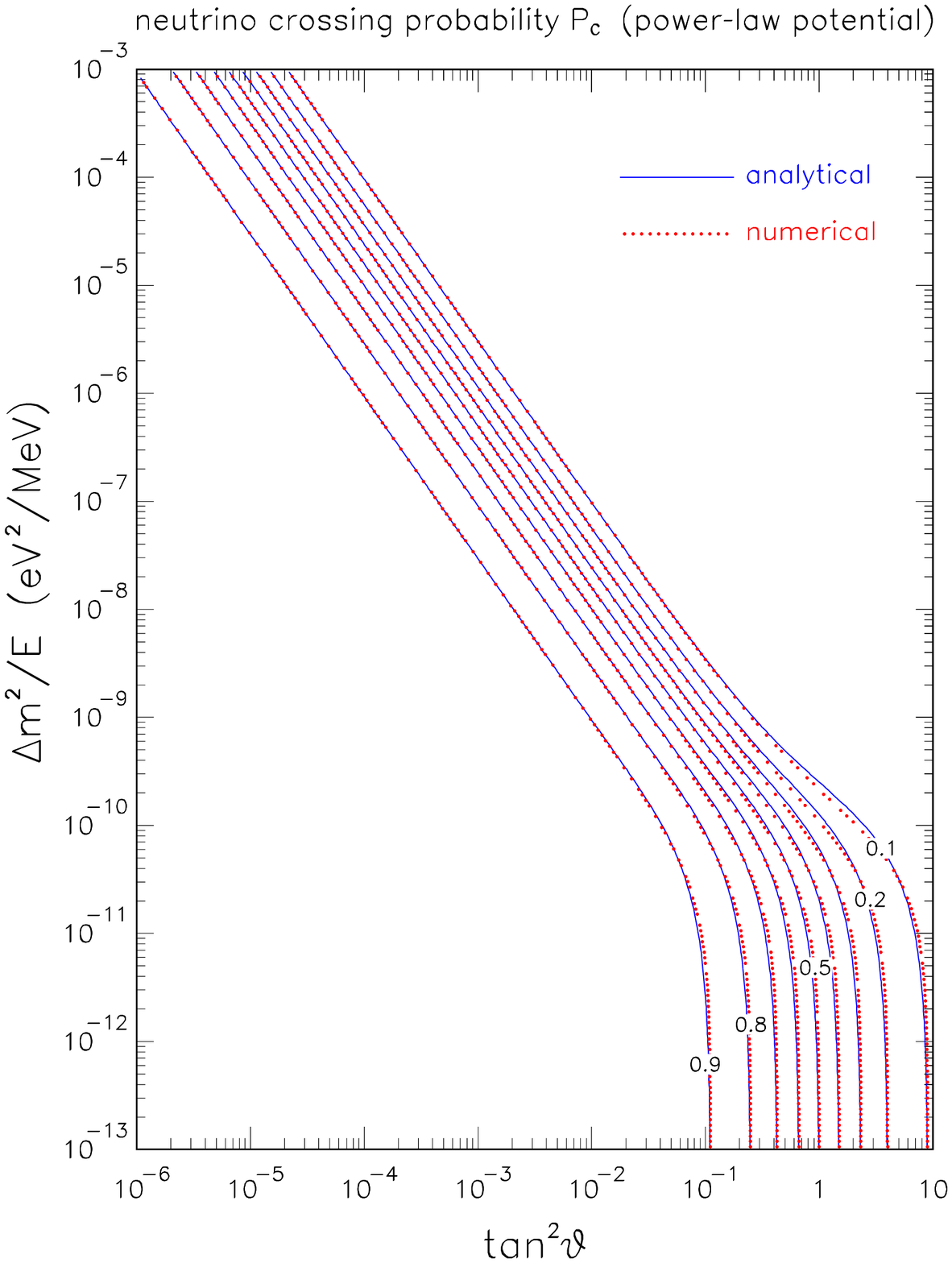}% 
{Fig.~2. Two-flavor transitions: neutrino crossing probability $P_c(\nu)$ in
the parameter space $(\Delta m^2/E,\,\tan^2\theta)$ for the power-law
potential profile in Fig.~1. Dotted curves:  exact numerical calculations.
Solid curves: results of the analytical prescription in
Eqs.~(\protect\ref{Vk})--(\protect\ref{Pc}). Isolines of $P_c$ for
antineutrinos (not shown) can be obtained by reflection around the axis
$\tan^2\theta=1$. See the text for details.}
%..............................................................................
\InsertFigure{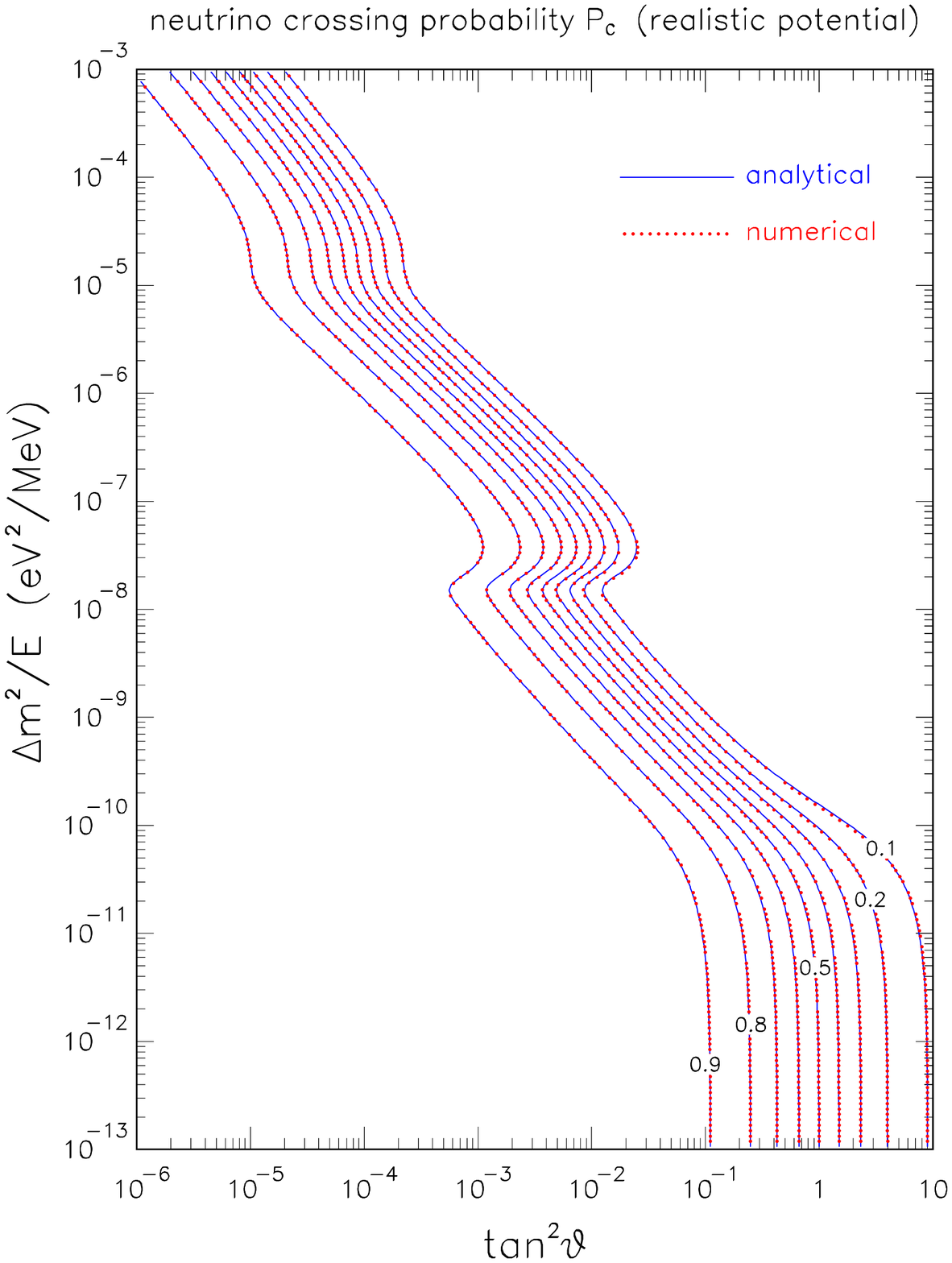}% 
{Fig.~3. As in Fig.~2, but for the realistic potential profile in  Fig.~1.}
%..............................................................................
\InsertFigure{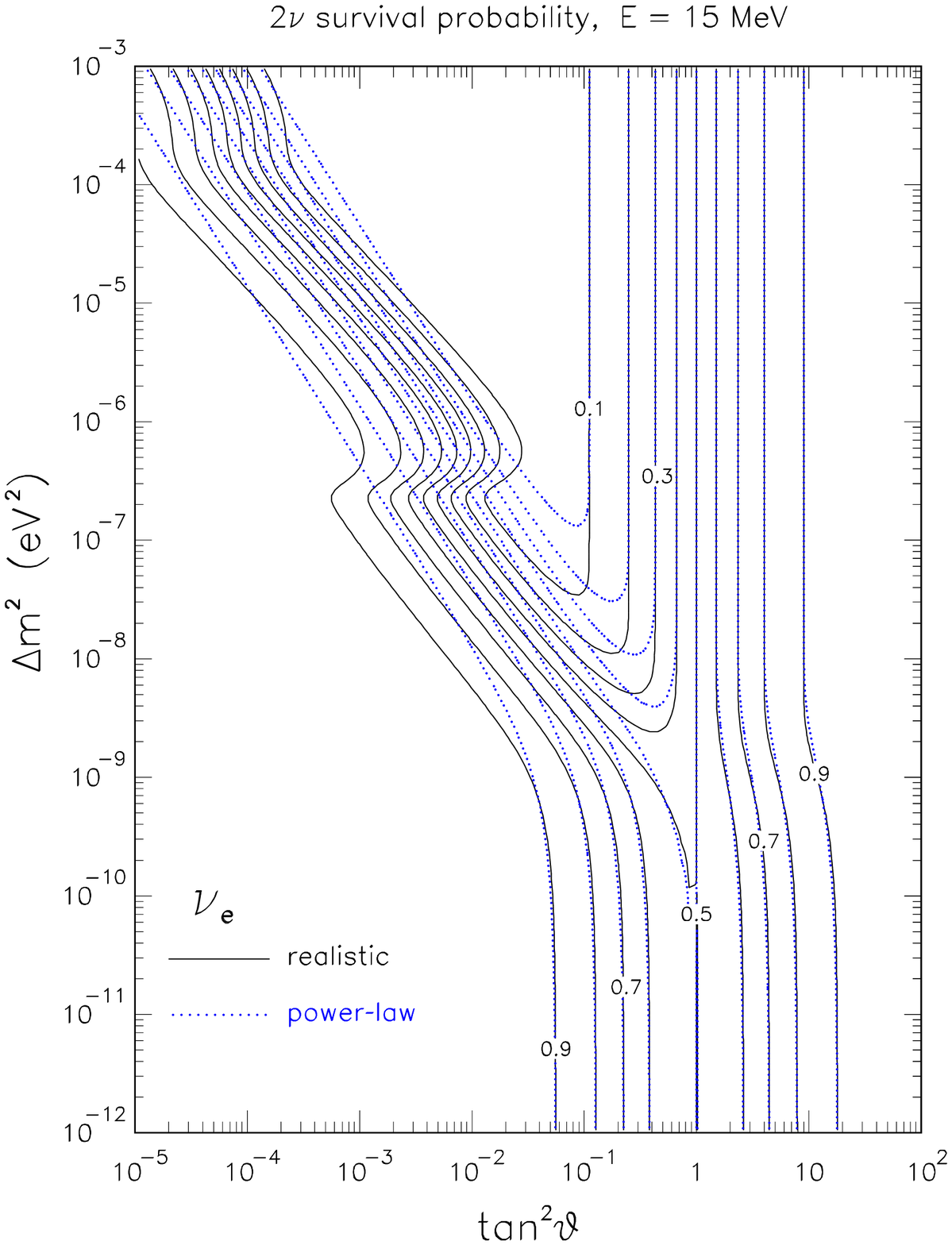}% 
{Fig.~4. Two-flavor transitions: analytical results for the electron neutrino
survival probability $P_{ee}$ in the mass-mixing plane $(\Delta
m^2,\,\tan^2\theta)$, at a representative neutrino energy ($E=15$~MeV).
Solid curves: realistic potential. Dotted curves: power-law potential.
Isolines of $P_{ee}$ for antineutrinos (not shown) can be obtained by
reflection around the axis $\tan^2\theta=1$. See the text for details.}
%..............................................................................
\InsertFigure{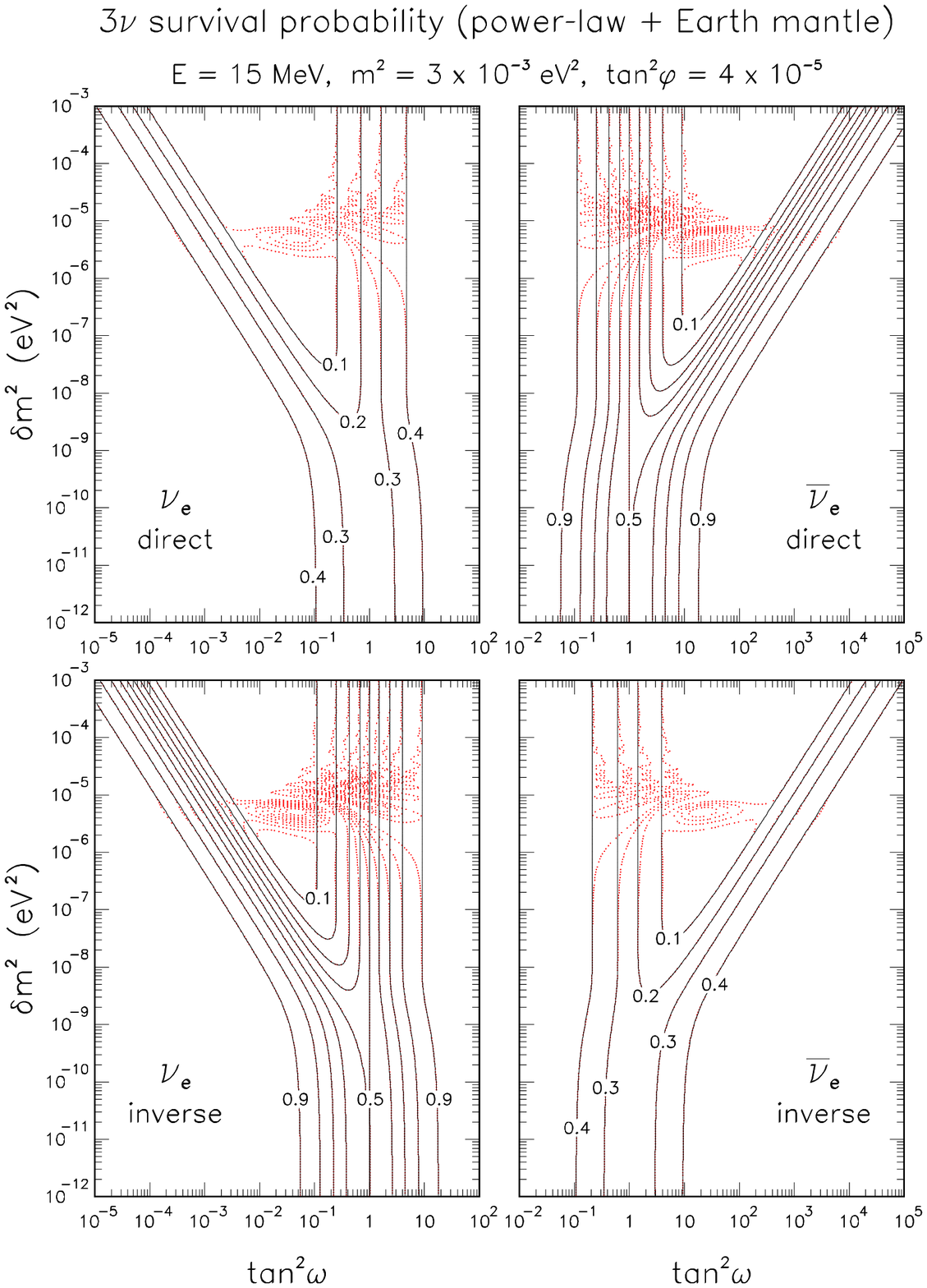}% 
{Fig.~5. Three-flavor transitions: analytical results for $P_{ee}$ in the
mass-mixing subspace $(\delta m^2,\,\tan^2\omega)$, assuming the power-law 
profile in Fig.~1 and fixing $E=15$~MeV, $m^2=3\times 10^{-3}$~eV$^2$, and
$\tan^2\phi=4\times 10^{-5}$ ($P_H^+=0.48$). Dotted lines include Earth matter
effects for a representative path of 8500 km in the mantle (assuming
$\rho=4.5$ g/cm$^3$ and $Y_e=1/2$). The upper panels refer to $\nu$ (left) and
$\overline\nu$ (right) in the case of direct  hierarchy. The lower panels
refer to $\nu$ (left) and $\overline\nu$ (right) in the case of inverse
hierarchy.}
%..............................................................................

\eject
\end{document}